\documentclass[runningheads]{llncs}

\usepackage[utf8]{inputenc}
\usepackage{graphicx}
\usepackage{hyperref}
\usepackage{amsmath}

\title{An attempt at beating the 3D U-Net}

\author{
Fabian Isensee\inst{1,2} \and
Klaus H. Maier-Hein \inst{1}
}
\authorrunning{F. Isensee and Klaus H. Maier-Hein}
\institute{Division of Medical Image Computing, German Cancer Research Center (DKFZ), Heidelberg, Germany \and
Faculty of Biosciences, University of Heidelberg, Heidelberg, Germany}
\begin{document}

\maketitle  

\begin{abstract}
The U-Net is arguably the most successful segmentation architecture in the medical domain. Here we apply a 3D U-Net to the 2019 Kidney and Kidney Tumor Segmentation Challenge and attempt to improve upon it by augmenting it with residual and pre-activation residual blocks. Cross-validation results on the training cases suggest only very minor, barely measurable improvements. Due to marginally higher dice scores, the residual 3D U-Net is chosen for test set prediction. With a Composite Dice score of 91.23 on the test set, our method outperformed all 105 competing teams and won the KiTS2019 challenge by a small margin.
\end{abstract}

\section{Introduction}
With over 400,000 cases per year, kidney tumors pose a serious health issue. Right now, surgery is the most common treatment option. Semantic segmentation of the kidneys and the tumorous tissue is a promising first step towards improving the treatment outcome, for example by serving as processing step in surgery planning \cite{taha2018kid}, or by enabling research that attempts to relate tumor morphology to surgical outcome \cite{ficarra2009preoperative,kutikov2009renal}. 

Fueled by the availability of publicly available databases, semantic segmentation is one of the most popular research topics in the medical image computing domain. This is especially true for abdominal CT scans, where several competitions encourage researchers to continue developing methods to increase their segmentation performance \cite{litsPaper,BCVAbdomenChallenge,decathlonDataPaper}. Despite the broad availability of abdominal CT data, some of which also includes the kidneys as segmentation targets, no public dataset with kidney tumor labels has been available until now. This naturally also results in a relatively low number of segmentation algorithms that are specifically designed to segment kidney tumours \cite{yu2018crossbar,skalski2016kidney,yang2014automatic}. The kidney tumor segmentation challenge (KiTS) \cite{heller2019kits19} aims at tackling this deficiency by providing 210 high quality annotated CT scans for training and 90 CT scans for algorithm testing. 

The vast majority of successful algorithms for 3D image segmentation in the medical domain is based on 3D variants of the U-Net architecture \cite{ronneberger2015u,cciccek20163d}. While the U-Net is thereby commonly augmented using residual \cite{milletari2016v} or dense \cite{li2018h} connections, recent work has achieved excellent results using just a plain U-Net architecture \cite{isensee2019nnu}, questioning the necessity of extensive architecture research in the medical domain.

This paper picks up on the success of the U-Net. While it is heavily based on \cite{isensee2019nnu}, we here attempt to revise the strong abstinence regarding architectural modification by employing different U-Net-inspired models, some of which make use of residual blocks in the encoder. All experiments are done in the context of the KiTS 2019 challenge and we use the best U-Net model (based on cross-validation on the training set) for our challenge submission.

\section{Method}
Based on the success of the U-Net architecture, we develop and train three different U-Net inspired architectures: a 3D 'plain' U-Net (no residual/dense connections), a residual \cite{he2016deep} 3D U-Net and a pre-activation \cite{he2016identity} residual 3D U-Net. 

\subsection{Preprocessing}
It is common in large datasets such as the KiTS dataset that voxel spacings are inhomogeneous. Convolutional neural networks cannot interpret voxel spacings natively, which is why we preprocess the KiTS dataset by resampling all cases to a common voxel spacing (called target spacing). Due to the amount of available GPU memory, the patch size that can be processed in 3D CNNs is typically quite limited. Thus, the target spacing, which directly impacts the total size of the images in voxels, also determines how much contextual information the CNN can capture in its patch size. Conversely, increasing the voxel spacing too much will reduce the image size to a point where detailed information to, for example, properly distinguish cysts from tumors may be lost. Optimizing the trade-off between the amount of contextual information in the networks patch size vs the details retained in the image data is crucial in obtaining ideal performance. Here we resample all cases to a common voxel spacing of $3.22\times 1.62\times 1.62$ mm, resulting in a median image shape of $128\times 248\times 248$ voxels for the training cases. 

CT images are quantitative. Thus, the same intensity values are expected to be identical when examining the same organ on scans originating from different scanners or hospitals. This property is often exploited to set a \textit{level window}, where intensities are clipped to some organ-specific value range. We adopt this idea for our preprocessing: each case is clipped to the range $[-79, 304]$. We then subtract $101$ and divide by $76.9$ to bring the intensity values in a range that is more easily processed by CNNs (due to the nature of weight initialization).

\subsection{Network architecture}
As stated previously, we employ three 3D U-Net architectures for our experiments. All U-Nets use 3D convolutions, ReLU/LReLU nonlinearities and instance normalization. Upsampling is done via transposed convolution and downsampling is done with strided convolutions. All networks start with some number of feature maps at the highest resolution. This number is is doubled with each downsampling operation (up to a maximum of 320 feature maps) in the encoder and halved with each transposed convolution in the decoder. We always downsample by a factor of 2. Downsampling is done until further downsampling would result in a spatial feature map size $<$ 4.

\begin{figure}[t]
    \centering
    \includegraphics[width=0.8\textwidth]{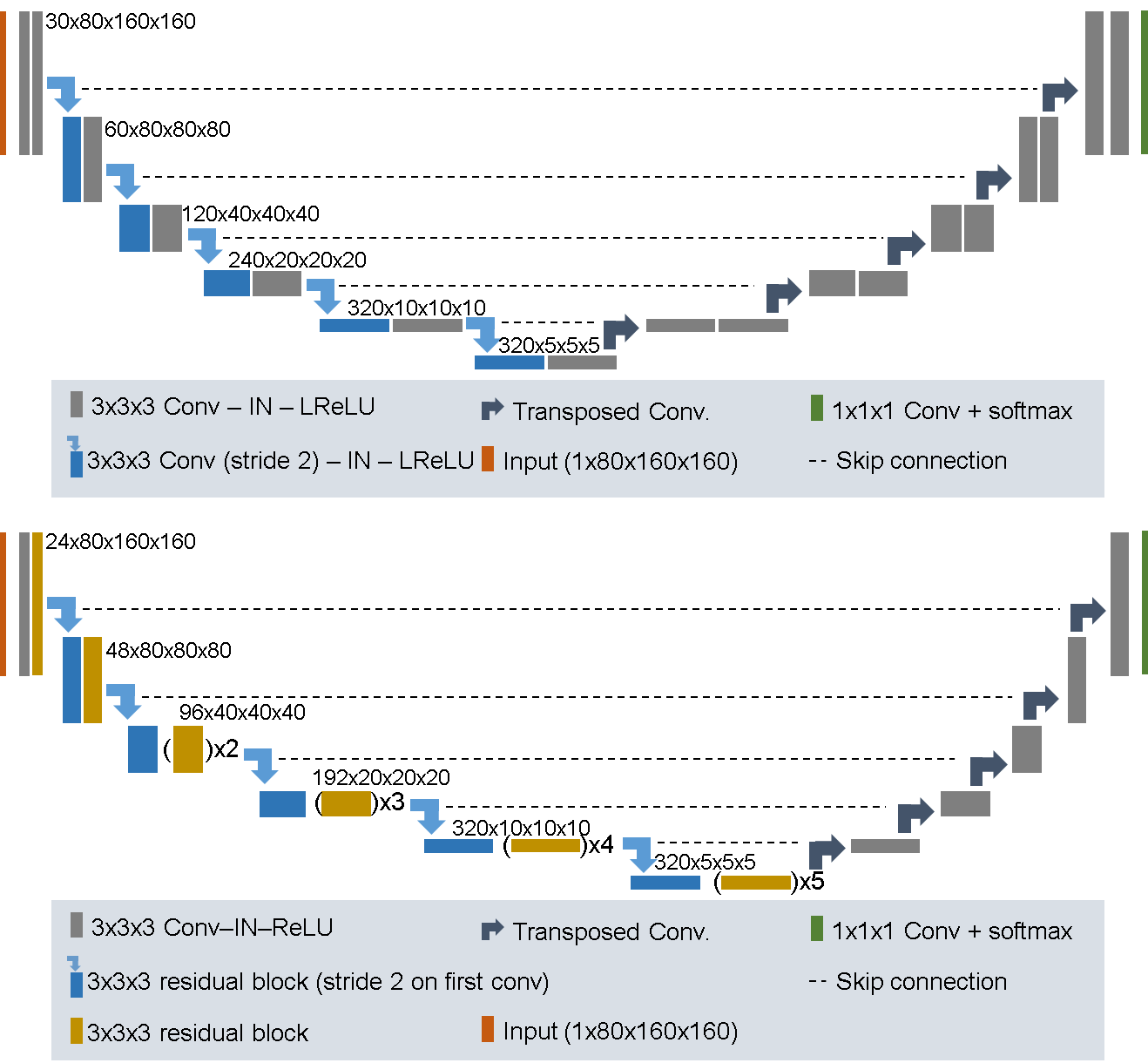}
    \caption{3D U-Net (top) and residual 3D U-Net architecture (bottom) used in this project. $(  )\times X$ denotes that a block is repeated X times. The architecture of the pre-activation residual U-Net is analogous to the residual U-Net (with instnorm and ReLU being shifted to accommodate pre-activation residual blocks).
    }
    \label{fig:architectures}
\end{figure}

\paragraph{Plain 3D U-Net}
For both the encoder and decoder we use two conv-instnorm-LReLU blocks between poolings/upsamplings. This architecture uses 30 feature maps at the highest resolution.

\paragraph{Residual 3D U-Net}
This architecture uses residual blocks in the encoder as opposed to a simple sequence of convolutions. The residual blocks are implemented similar to \cite{he2016deep}: conv-instnorm-ReLU-conv-instnorm-ReLU (where the addition of the residual takes place before the last ReLU activation). We start with just one residual block at the highest resolution and increase the number of residual blocks after each downsampling operation. The decoder uses only one conv-instnorm-ReLU per resolution. To accommodate the larger memory footprint of residual networks, we reduce the initial number of feature maps from 30 to 24.

\paragraph{Pre-activation residual 3D U-Net}
Inspired by \cite{he2016identity} we also use a variant of the residual 3D U-Net that uses pre-activation residual blocks: instnorm-ReLU-conv-instnorm-ReLU-conv.

An overview over the architectures used is provided in Figure \ref{fig:architectures}.

\subsection{Network training}
All networks are trained with stochastic gradient descent and a batch size of 2. We found that a patch size of $80\times 160 \times 160$ yields sufficient contextual information while retaining necessary fine grained image information. We define an epoch as iteration over 250 batches and train for a total of 1000 epochs. The sum of cross-entropy and dice loss is used as training objective and we use supervision at different resolutions to encourage gradient flows deeper into the network. The training of a single network utilizes 12 GB of VRAM and runs for about 5 days. Training was done on Nvidia Titan Xp GPUs (single GPU training). All networks were implemented with the PyTorch framework \cite{paszke2017automatic} (version 1.1). We base our implementation on nnU-Net\footnote{\href{https://github.com/MIC-DKFZ/nnUNet}{https://github.com/MIC-DKFZ/nnUNet}} \cite{isensee2019nnu}.

During training, we apply extensive data augmentation to all training patches with the \textit{batchgenerators} framework \footnote{\href{https://github.com/MIC-DKFZ/batchgenerators/}{https://github.com/MIC-DKFZ/batchgenerators/}}. We make use of scaling, rotations, brightness, contrast, gamma and Gaussian noise augmentations.

\subsection{Dataset modifications}
\label{modifications}
As was pointed out in \footnote{\href{https://github.com/neheller/kits19/issues/21}{https://github.com/neheller/kits19/issues/21}}, some cases in the training set may have been mislabeled. We manually inspected the cases with the worst tumor dice in our cross-validations and based on these evaluations made the following changes to the dataset:

\begin{enumerate}
    \item Case IDs 23, 68, 125 and 133 were excluded because our networks were in disagreement with the provided reference annotation and we felt we had insufficient expertise to decide on the correctness of either segmentation.
    \item Case IDs 15 and 37 were confirmed to be faulty by the challenge organizers. Therefore we replaced their reference annotation with the segmentation generated by previous iterations of our networks (visual inspection seemed plausible, no manual fine tuning of the segmentations was performed).
\end{enumerate}

Note that excluding and modifying training cases was explicitly permitted by the challenge organizers\footnote{https://discourse.kits-challenge.org/t/rules-automatic-or-manual-modification-to-the-training-data/101/2}.

\section{Results}
Dice scores for \textit{kidney} were computed by treating both the actual kidney label as well as the tumor label as foreground and everything else as background. This constitutes the same setup that is used in the challenge evaluation. The dice computation of the tumors is done simply on the tumor labels. No other metrics are considered as the challenge is evaluated on the geometric mean of kidney and tumor dice.

All scores shown in this section are based on five-fold cross-validations on the KiTS training dataset.

\begin{table}[]
    \centering
    \caption{Summary of experiments. Results are based on five-fold cross-validations on the training cases of the KiTS dataset. Note that our training dataset was slightly modified. Thus, the cross-validation Dice scores are not necessarily comparable with other challenge submissions.}
    \begin{tabular}{l|c|c|c}
    Network architecture  & Kidney Dice & Tumor Dice & Composite Dice \\ \hline
    3D U-Net              & 97.34       & 85.04      & 91.19     \\
    Residual 3D U-Net     & 97.36       & \textbf{85.73}      & \textbf{91.54}     \\
    Preact. Res. 3D U-Net & 97.37       & 85.13      & 91.25     \\
    ensemble              & \textbf{97.43}       & 85.58      & 91.50    
    \end{tabular}
    \label{tab:results}
\end{table}

The challenge is decided based on the mean Dice score between kidney and tumor ('Composite Dice'). In case of a draw between two teams, the tumor dice is used as a tie breaker.
As can be seen in Table \ref{tab:results}, the results between the three 3D U-Net models are remarkably similar, even to a point where we don't feel comfortable to declare one of the models better than the others. This is particularly interesting given the amount of publications that claim substantial performance improvements when altering the U-Net architecture.

Even ensembling, which is typically a go-to technique to improve the segmentation quality over single models, did not yield an improvement. This makes the decision of what model should be submitted for the test cases particularly challenging. 

Based on the cross-validation results, the Residual 3D U-Net seems most promising, as it has the highest mean dice and tumor dice scores (the metrics that matter in the context of this challenge). Still, we would like emphasize that this decision is being made out of necessity rather than conviction.

We thus use the five Residual 3D U-Net models from the cross-validation as an ensemble to predict the 90 test cases. Predictions were consolidated by averaging softmax outputs.

\begin{table}[]
    \centering
    \caption{Test set results on KiTS 2019 (\href{http://results.kits-challenge.org/miccai2019}{http://results.kits-challenge.org/miccai2019}). Teams were ranked by Composite Dice. Bold numbers indicate the highest scores across all participating teams. This table shows the first five teams only, please refer to the online leaderboard for the scores of all participants.}
    \begin{tabular}{l|c|c|c}
    Team (Rank)  & Kidney Dice & Tumor Dice & Composite Dice \\ \hline
    Isensee F et al. (\textbf{ours}) (1)             & 97.37       & \textbf{85.09}      & \textbf{91.23}     \\
    Xiaoshuai Hou et al. (2)     & 96.74       & 84.54      & 90.64     \\
    Guangrui Mu et al. (3)     & 97.29       & 83.21      & 90.25     \\
    Yao Zhang et al. (4) & 97.42       & 83.06      & 90.24     \\
    Jun Ma (5)  & 97.34       & 82.54      & 89.94   
    \end{tabular}
    \label{tab:results_test}
\end{table}

Table \ref{tab:results_test} shows the top 5 results on the KiTS 2019 test set. Our method outperformed all competing methods and thus won the KiTS 2019 challenge.

\section{Discussion}
With a Composite Dice score of 91.23 on the test set, our method outperformed all 105 competing teams and won the KiTS2019 challenge by a small margin (0.6 Composite Dice to the second place).

Our comparison between a simple U-Net architecture on the one hand and residual U-Net architectures on the other yielded inconclusive results: we were unable to declare a clear winner based on our experiments. Even ensembling our three models did not yield improvements over the best single-model result. Still, for the sake of this challenge, something needed to be selected for test set prediction. Due to marginally higher scores, the residual 3D U-Net was chosen for this purpose. 

It is interesting to see that architectural modifications do not significantly improve segmentation results here, especially because more advanced architectural designs are typically believed to also improve the segmentation results. We should point out, however, that the lack of improvement could be due to a number of reasons: bad implementation of the residual networks, bad choice of hyperparameters for some of the architectures, etc. To give a definitive answer, a much more thorough study with extensive hyperparameter optimization for each of the architectures is warranted. We would also like to emphasize that the conclusions we drew from our experimental comparison in this paper were not tested for statistical significance. 

Note that while the 3D U-Net used leaky ReLU nonlinearities (negative slope $10^{-2}$) while its residual counterparts used ReLUs. This is due to an implementation error in the network architecture of the 3D U-Net. We do not expect this to impact the results and thus the conclusions of this work.

It is furthermore important to point out that we base our comparison between architectures not on keeping a constant number of parameters or layers but instead choose a much more realistic constraint: What is the best we can do with an architecture given that it needs to fit some hardware constraint (12 GB Titan X GPU)? We believe this to be a much more sensible and reasonable constraint, especially in the context of challenges, because the amount of available GPU memory is typically more limiting than the number of multiply-add operations or the size of saved models. 

Generally, when designing an algorithm for a challenge, we like to implement a baseline and then improve upon it. Here, however, we were unable to substantially improve upon the baseline, our 3D U-Net architecture. Several additional experiments were done in the context of this challenge that are not shown in this manuscript for brevity. For example, among other things, we attempted to vary the target spacing for resampling as well as the patch size of our network architecture. Usually, this is one of the most important hyperparameters that needs to be adjusted for a new dataset. Across all tested configurations (which were definitely biased towards what we deemed reasonable), the results were basically identical. To us this indicates that the segmentation problem posed in this dataset is well-behaved: it can be solved effectively with relatively simple baseline methods and is quite robust with respect to the choice of hyperparameters. We therefore expected the competition in this challenge to be exceptionally strong and, if it wasn't for the slightly decreased tumor Dice score of the ensemble, we would have opted for using this over just the Residual 3D U-Net to gain a competitive edge over other teams.

While this work was designed to achieve maximum segmentation accuracy, future work should be directed towards more extensively evaluating potential differences between the 'plain' U-Net and its architectural variants. Such a comparison should include extensive hyperparameter optimization for each of the architectures and also make a thorough statistical analysis of the results.

\bibliographystyle{splncs04}
\bibliography{bibliography}
 
\end{document}